\documentclass[printer]{aa}

\usepackage{epsfig}
\usepackage{graphics}
\usepackage[latin1]{inputenc}
\usepackage{natbib}

\usepackage{amssymb}
\usepackage{amsmath}

\defcitealias{bib:neveu}{N13} 	
\defcitealias{bib:barreira13}{B13a} 	

\begin{document}
\title {First experimental constraints on the\\disformally coupled Galileon model}
\author{J.~Neveu\inst{1}, V.~Ruhlmann-Kleider\inst{1}, P.~Astier\inst{2},
M.~Besan\c{c}on\inst{1}, A.~Conley\inst{3}, 
J.~Guy\inst{2}, A.~M\"oller\inst{1}, N.~Palanque-Delabrouille\inst{1},
E. Babichev\inst{4}
}
\institute{CEA, Centre de Saclay, Irfu/SPP,  91191 Gif-sur-Yvette, France
\and  LPNHE, Universit\'e Pierre et Marie Curie, Universit\'e Paris Diderot, CNRS-IN2P3, 4 place Jussieu, 75252 Paris Cedex 05, France
\and Center for Astrophysics and Space Astronomy, University of Colorado, Boulder, CO 80309-0389, USA
\and Laboratoire de Physique Th\'eorique d'Orsay, B\^atiment 210,
Universit\'e Paris-Sud 11, F-91405 Orsay Cedex, France
%\and $\mathcal{G}\mathbb{R}\varepsilon\mathbb{C}\mathcal{O}$, Institut d'Astrophysique de Paris, UMR 7095-CNRS, Universit\'e Pierre et Marie Curie-Paris 6, 98bis boulevard Arago, F-75014 Paris, France
}

\date{\today}
\authorrunning{J. Neveu et al.}
\titlerunning{First experimental constraints on the disformally coupled Galileon model}

%\abstract{}{The Galileon model has been shown to be a good candidate to explain the late-time accelerated expansion of the Universe. In addition to good theoretical properties, it modifies gravity in such a way that it agrees with the recent cosmological data, and protects local General Relativity thanks to the Vainshtein screening effect. Moreover, this model can arise from braneworld constructions or massive gravity and in this case exhibits a disformal coupling to matter which can be studied. }{ In this paper, we updated the cosmological constraints on the uncoupled Galileon model published in a previous work, and compare them to other studies. We then derived the first experimental constraints on the Galileon model disformally-coupled to matter, using precise measurements of the cosmological distances and the growth rate of cosmic structures.}{In the uncoupled case, we observe a tiny improvement of the small tension observed between the constraints set by growth data and those from distances. In the coupled Galileon model, the best-fit scenario gives a better agreement with data, and favours at the $2.2\sigma$ level a non-zero disformal coupling to matter. This gives an interesting inlight on the possible extra-dimensional origin of the Galileon theory.}{}
\abstract{}{The Galileon model is a modified gravity model that can explain the late-time accelerated expansion of the Universe. In a previous work, we derived experimental constraints on the Galileon model with no explicit coupling to matter and showed that this model agrees with the most recent cosmological data. In the context of braneworld constructions or massive gravity, the Galileon model exhibits a disformal coupling to matter, which we study in this paper.}{After comparing our constraints on the uncoupled model with recent studies, we extend the analysis framework to the disformally coupled Galileon model and derive the first experimental constraints on that coupling, using precise measurements of cosmological distances and the growth rate of cosmic structures.}{In the uncoupled case, with updated data, we still observe a low tension between the constraints set by growth data and those from distances. In the disformally coupled Galileon model, we obtain better agreement with data and favour a non-zero disformal coupling to matter at the $2.5\sigma$ level. This gives an interesting hint of the possible braneworld origin of Galileon theory.}{}
\maketitle

\section{Introduction}

Dark energy remains one of the deepest mysteries of cosmology today. Even though it has been fifteen years since the discovery of dark energy \citep{bib:riess,bib:perlmutter}, its fundamental nature remains unknown. Adding a cosmological constant ($\Lambda$) to Einstein's general relativity is the simplest way to account for this observation, and leads to remarkable agreement with all cosmological data so far (see e.g. \cite{bib:planck}). However, the cosmological constant requires considerable fine-tuning to explain current observations and motivates the quest for alternative explanations of the nature of dark energy.

Modified gravity models aim to provide such an explanation. The Galileon theory, first proposed by \cite{bib:nicolis} involves a scalar field, hereafter called $\pi$, whose equation of motion must be of second order and invariant under a Galilean shift symmetry $\partial_\mu \pi \rightarrow \partial_\mu \pi + b_\mu$, where $b_\mu$ is a constant vector. This symmetry was first identified as an interesting property in the DGP model \citep{bib:dgp}. \cite{bib:nicolis} derived the five possible Lagrangian terms for the field $\pi$,  which were then formulated in a covariant formalism by \cite{bib:deffayet} and \cite{bib:deffayetb}.

This model forms a subclass of general tensor-scalar theories involving only up to second-order derivatives originally found by Horndeski \citep{bib:horndeski}. Later, Galileon theory was also found to be the non-relativistic limit of numerous broader theories, such as massive gravity \citep{bib:deRhamMasGra} or brane constructions \citep{bib:deRhamDBI,bib:hinterbichler,bib:acoleyen}. Braneworld approaches give a deeper theoretical basis to Galileon theories. The usual and simple construction involves a 3+1 dimensional brane, our Universe, embedded in a higher dimensional bulk. The Galileon field $\pi$ can be interpreted as the brane transverse position in the bulk, and the Galilean symmetry appears naturally as a remnant of the broken space-time symmetries of the bulk \citep{bib:hinterbichler}. The Galilean symmetry is then no longer imposed as a principle of construction, but is a consequence of space-time geometry.

Models that modify general relativity have to alter gravity only at cosmological scales in order to agree with the solar system tests of gravity (see e.g. \cite{bib:will}). %They usually introduce a new scalar field, mediator of a fifth force, which plays the role of dark energy at large scale but also modify gravity at short scales. Two mechanisms exist to screen that fifth force at small scales: the chameleon effect and the Vainshtein effect \citep{bib:vainshtein}. The first one mainly appears in models such that $f(R)$ theories (see e.g. \cite{bib:sotiriou,bib:felice2010fR}), while 
The Galileon field can be coupled to matter either explicitly or through a coupling induced by its temporal variation \citep{bib:babichev13}. This leads to a so-called fifth force that by definition modifies gravity around massive objects like the Sun. But the non-linear Lagrangians of the Galileon theory ensure that this fifth force is screened near massive objects in case of an explicit coupling of the form $c_0 \pi T^\mu_{\ \mu}/M_P$ (where $T^\mu_{\ \mu}$ is the trace of the matter energy-momentum tensor, $c_0$ a dimensionless parameter, and $M_P$ the Planck mass) or in the case of an induced coupling. This is called the Vainshtein effect (\cite{bib:vainshtein} and \cite{bib:babichev13b} for a modern introduction). The fifth force is thus negligible with respect to general relativity within a certain radius from a massive object, that depends on the object mass and Galileon parameters \citep{bib:vainshtein,bib:nicolis,bib:brax11}.

Braneworld constructions and massive gravity models give rise to an explicit disformal coupling to matter of the form $\sim \partial_\mu \pi \partial_\nu \pi T^{\mu\nu}$. As shown, for example, in \cite{bib:brax12b}, this coupling does not induce a fifth force on massive objects since it does not apply to non-relativistic objects when the scalar field is static. In a cosmological context, the scalar field $\pi$ evolves with time but the fifth force introduced by the disformal coupling to matter can be masked thanks to a new screening mechanism \citep{bib:koivisto,bib:zumalacarregui}. However, the disformal coupling still plays a role in the field cosmological evolution, which makes this kind of Galileon model interesting to compare with cosmological data. The action of the model is
\begin{equation}\label{eq:action}
S=\int d^4x \sqrt{-g}\left( \frac{M_P^2R}{2} - \frac{1}{2} \sum_{i=1}^{5} c_i L_i - L_m - L_G\right),
\end{equation}
with $L_m$ the matter Lagrangian, $R$ the Ricci scalar, and $g$ the determinant of the metric. 
The $c_i$s are the Galileon model dimensionless parameters weighting different covariant Galileon Lagrangians $L_i$ \citep{bib:deffayet}:
%(\nabla_\mu \pi\nabla^\mu \pi)
\begin{align}
L_1 = & M^3\pi,\quad L_2=\pi_{;\mu}\pi^{;\mu},\quad
L_3=(\pi_{;\mu}\pi^{;\mu})(\square \pi)/M^3, \notag \\
L_4 = & (\pi_{;\mu}\pi^{;\mu})\left[ 2(\square \pi)^2 - 2 \pi_{;\mu\nu}\pi^{;\mu\nu} - R(\pi_{;\mu}\pi^{;\mu})/2 \right]/M^6, \notag \\
L_5 = & (\pi_{;\mu}\pi^{;\mu})
\left[ (\square \pi)^3 - 3(\square \pi) \pi_{;\mu\nu}\pi^{;\mu\nu} +2\pi_{;\mu}\ ^{;\nu}\pi_{;\nu}^{\ ;\rho}\pi_{;\rho}^{\ ;\mu} \right. \notag \\ & \left. -6 \pi_{;\mu}\pi^{;\mu\nu}\pi^{;\rho}G_{\nu\rho}\right]/M^9,
\end{align}
where $M$ is a mass parameter defined as $M^3=H_0^2M_P$ with $H_0$ the Hubble parameter current value. $L_G$ is the disformal coupling to matter:
\begin{equation}
L_G=\frac{c_G}{M_P M^3}\partial_\mu \pi \partial_\nu \pi T^{\mu\nu},
\end{equation}
where $c_G$ is dimensionless. Interestingly, \cite{bib:babichev} showed that $c_0\lesssim 10^{-2}$ by comparing local time variation measurements of the Newton constant $G_N$ in the Lunar Laser Ranging experiments, to predictions derived in the Galileon theory with the Vainshtein mechanism accounted for and boundary conditions set by the cosmological evolution. %, while the induced coupling is of order of unity \citep{bib:babichev13}. 
In the more general context of scalar field theories, the disformal coupling has been recently constrained in particle physics using Large Hadron Collider data \citep{bib:brax14,bib:monophoton}.
Thus the disformal coupling should be the first explicit Galileon coupling to look at considering the actual existing constraints.

The uncoupled Galileon model ($c_G=0$) has already been constrained by observational cosmological data in \citet{bib:appleby2}, \citet{bib:okada}, \citet{bib:nesseris}, and more recently in \cite{bib:neveu} (hereafter \citetalias{bib:neveu}) and \cite{bib:barreira13} (hereafter \citetalias{bib:barreira13}). In \citetalias{bib:neveu}, we introduced a new parametrisation of the model and developed a likelihood analysis method to constrain the Galileon parameters independently of initial conditions on the $\pi$ field. The unknown initial condition for the Galileon field was absorbed into the original $c_i$ parameters to form new parameters $\bar c_i$ defined by $\bar c_i = c_i x_0^i$, where $x_0$ encodes the initial condition for the Galileon field. The same methodology was adopted here, and we refer the interested reader to \citetalias{bib:neveu} for more details. 

Same datasets were used %for high-quality type Ia supernovae (SNe Ia) \citep{bib:guy2010,bib:conley,bib:sullivan}, 
for baryonic acoustic oscillations (BAO) \citep{bib:beutler,bib:padmanabhan,bib:sanchez}, and for growth of structure joint measurements\footnote{In order to ensure that the measurements do not depend of a fiducial cosmology.} of $f\sigma_8(z)$ and the Alcock-Paczynski parameter $F(z)$, mainly from \cite{bib:percival04,bib:blake11b,bib:beutler12,
bib:samushia12a,bib:reid}. For the cosmic microwave background (CMB), we updated our analysis 
  to use the WMAP9 distance priors \citep{bib:wmap9} instead of the 
  WMAP7 ones \citep{bib:komatsu11}. Concerning type Ia supernovae (SNe Ia), we also updated our sample from the high-quality data of the SuperNova Legacy Survey (SNLS) collaboration  \citep{bib:guy2010,bib:conley,bib:sullivan} to the recent sample published jointly by the SNLS and Sloan Digital Sky Survey (SDSS) collaborations \citep{bib:jla}, which we will refer to as the Joint Light-curve Analysis (JLA) sample in the following. 

Interesting constraints on the uncoupled Galileon model using the full CMB power spectrum were published in \citetalias{bib:barreira13}, with a different methodology. In the following, we show that the results from both studies agree. However, in our study we used growth data, despite our using a
  linearised version of the theory, while \citetalias{bib:barreira13} preferred not to use those data until the Galileon non-linearities responsible for the Vainshtein effect are precisely studied. We include a discussion on that important point in this paper.% Their results using N-body simulations show the difficulty to model correctly that screening effect around massive objects (see \cite{bib:barreira13b,bib:li13,bib:barreira13c}). We therefore allow us in this work to continue using a linearised Galileon model, waiting for more conclusive studies. 

%With the updated data, we find that the uncoupled Galileon model is still not significantly disfavoured by current data, and that the tension between cosmological distances and growth data decrease. Interesting constraints on the disformally-coupled Galileon model are also provided, showing that data prefer a non-zero value for that coupling at the $2.2\sigma$ level, which can give clues on the braneworld origin of the Galileon model.

Section~\ref{sec:data} describes our updated datasets. Section~\ref{sec:uncoupled} provides an update of the constraints on the uncoupled Galileon model, using WMAP9 and JLA data, and a comparison with \citetalias{bib:barreira13} results. Section~\ref{sec:coupled} gives constraints on the disformally coupled Galileon model derived from the same dataset. Section~\ref{sec:disc} discusses these results and their implications, as well as the state of the art of growth rate of structure modelling in Galileon theory. We conclude in Section~\ref{sec:concl}.

\section{Updated datasets}\label{sec:data}

%In this work, we followed the same methodology as in \citetalias{bib:neveu}, with changes described hereafter. 

\subsection{Updated CMB data}

\begin{table}[h!]
\caption[]{WMAP distance priors.}
\label{tab:wmap}
\begin{center}
\begin{tabular}{ccc} \hline \hline \\ [-1ex]
   & WMAP7 & WMAP9 \\  [1ex] \hline \\ [-1ex]
$R$  & $1.725\pm0.018$ & $1.725\pm0.018$ \\ [1ex]  \hline \\ [-1ex]
$l_a$  & $302.09\pm0.76$  & $302.40\pm0.69$ \\ [1ex]  \hline \\ [-1ex]
$z_*$  & $1091.3\pm0.91$ & $1090.88\pm1.00$ \\ [1ex]  \hline \\ [-1ex]
\end{tabular}
\tablefoot{The uncertainties are computed using all terms of the inverse covariance matrices published in \cite{bib:komatsu11} and \cite{bib:wmap9}.}
\end{center}
\end{table}

The new CMB distance priors and their covariance matrix from WMAP9 (\cite{bib:wmap9}) were used as CMB constraints (see Table~\ref{tab:wmap}). No major improvements are expected as the distance priors uncertainties are not significantly decreased in the WMAP9 release. Planck results would be very competitive but the Planck collaboration did not publish similar distance priors independent of the $\Lambda$CDM model\footnote{Note however that recently an independent group proposed a derivation of the distance priors from the Planck+WP+lensing power spectrum \citep{bib:wang}.}.

In \citetalias{bib:neveu} as well as in this work, we followed the prescriptions of \cite{bib:komatsu09} for the use of WMAP distance priors to derive cosmological constraints. In particular, they recommend a minimisation procedure of the $h$ value when comparing predictions to observables. Because of the rich Galileon phase space to explore, we added in \citetalias{bib:neveu} a Gaussian prior on $H_0$ to help the program converge, centred on the \cite{bib:riess11} measurement. Since then, the Planck collaboration published their results and showed a disagreement between their $\Lambda$CDM fit and the \cite{bib:riess11} measurement for $H_0$. In order to measure the impact of that measurement on our results, we repeated %the fit increasing the $H_0$ uncertainty by a factor five
the study without the $H_0$ prior. Results on e.g. $\Omega_m^0$ and $\bar c_2$ are presented in Table~\ref{tab:h0}. When removing the $H_0$ prior, best-fit values and the minimised $h$ value do not change drastically. %$\Omega_m^0$ and $h$ best-fit values shifted in the expected directions. 
The $\chi^2$ at the marginalised values decreased from 2.1 to 0.7, indicating that there is a small tension between the WMAP9 distance priors and $H_0$ measurement. The check was also performed in the disformally coupled Galileon and $\Lambda$CDM models, and led to similar results. As a consequence, we decided to remove the $H_0$ prior in the following.

\begin{table}[htb]
%\caption[]{Impact of the $H_0$ Gaussian prior on the Galileon constraints using BAO+WMAP9+H0 data (top) and all data (bottom) in the coupled Galileon model.}
\caption[]{Impact of the $H_0$ Gaussian prior on constraints on two of the uncoupled Galileon model parameters using BAO+WMAP9 data.}
\label{tab:h0}
\begin{center}
\begin{tabular}{ccccc} \hline \hline \\ [-1ex]
 $H_0$ (km/s/Mpc)  & $\Omega^0_m$ & $\bar c_2$ & $h$ & $\chi^2$  \\  [1ex] \hline \\ [-1ex]
$73.8\pm2.4$  & $0.270^{+0.014}_{-0.009}$ & $-5.614^{+1.970}_{-2.650}$ & $0.714$ & 2.1 \\ [1ex]  \hline \\ [-1ex]
No prior  & $0.274^{+0.015}_{-0.009}$ & $-5.467^{+1.962}_{-2.659}$ & $0.704$ & 0.7 \\ [1ex]  \hline \\ [-1ex]
\end{tabular}
%\tablefoot{In the second line, the $H_0$ uncertainty is artificially increased by a factor five to measure the impact the $H_0$ prior on our constraints.}
\end{center}
\end{table}
\vspace*{-1cm}

\subsection{Updated SN Ia sample}

\begin{figure}[hbtp]
\begin{center}
\epsfig{figure=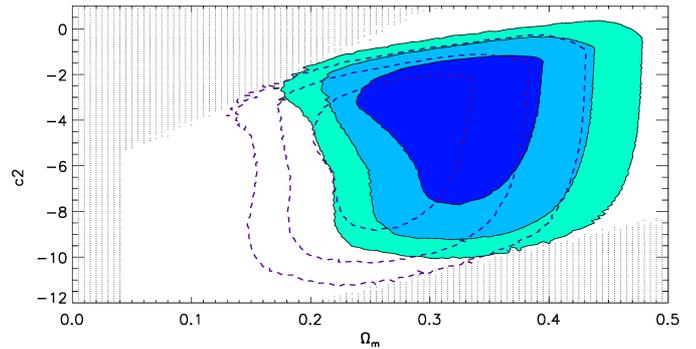, width=\columnwidth} 
\caption[]{Experimental constraints on $\Omega_m^0$ and $\bar c_2$ parameters of the uncoupled Galileon model from SNLS3 data (dashed purple contours) and JLA data (filled blue contours). $\alpha$ and $\beta$ were fixed to their $\Lambda$CDM best-fit values, and we marginalised over the remaining fitted parameters. The filled dark, medium, and light-blue contours enclose 68.3, 95.4, and 99.7\% of the probability, respectively. The contours include statistical and all identified systematic uncertainties. The dark dotted regions correspond to scenarios rejected by theoretical constraints.} 
\label{fig:snls}
\end{center}
\vspace*{-1cm}
\end{figure}

\begin{table*}[htbp]
\caption[]{Cosmological constraints on the uncoupled Galileon model from the SNLS3 and JLA samples.}
\label{tab:snls}
\begin{center}
\begin{tabular}{cccccccccc} \hline \hline \\ [-1ex]
Sample & $\Omega_m^0$ & $\bar c_2$ & $\bar c_3$ & $\bar c_4$ & $\alpha$ & $\beta$ & $\mathcal{M}_B^1$ & $\mathcal{M}_B^2$ & $\chi^2$ \\  [1ex] \hline \\ [-1ex]
SNLS3 Stat+sys & $0.273^{+0.054}_{-0.042}$ & $-5.240^{+1.880}_{-2.802}$ & $-1.781^{+1.071}_{-1.426}$ & $-0.588^{+0.516}_{-0.348}$ & 1.428 & 3.263 & 23.997 & 23.950 & 415.4 \\ [1ex] \hline \\ [-1ex] 
JLA Stat+sys & $0.328^{+0.055}_{-0.047}$ & $-4.175^{+1.726}_{-3.027}$ & $-1.345^{+0.968}_{-1.542}$ & $-0.475^{+0.464}_{-0.349}$ & 0.141 & 3.101 & 24.072 & 24.081 & 692.8\\ [1ex]  \hline \\ [-1ex] 
\end{tabular}
\tablefoot{Results were computed using statistical and systematic uncertainties combined. $\alpha$ and $\beta$ were kept fixed to their marginalised values. No errors are given on $\mathcal{M}_B^1$ and $\mathcal{M}_B^2$ because they were analytically marginalised over (see \cite{bib:conley}).}
\end{center}
\end{table*}

In \citetalias{bib:neveu}, the SNLS3 sample from \cite{bib:conley} was used. The JLA sample recently released by the SNLS and SDSS collaborations benefit from reduced calibration systematic uncertainties and combine the full SDSS-II spectroscopically-confirmed SN Ia sample with the SNLS3 sample. While the SNLS3 sample contain 472 supernovae whose parameters were determined using a combination of the SALT2 and SiFTO light-curve fitters, 740 supernovae are present in the final JLA sample, measured using SALT2 only. The impact of the new calibration and change in the light-curve fitter shifted the best fit $\Omega_m^0$ value for a flat $\Lambda$CDM model from $0.228 \pm 0.038$ to $0.295\pm0.034$ ($1.8\sigma$ drift, see \cite{bib:betoule13} and section 6.4 of \cite{bib:jla} for more details). This new value is now more consistent with the Planck measurement \citep{bib:planck}.

Both samples were used to derive constraints on the Galileon model parameters. Results are presented in Table~\ref{tab:snls} and illustrated in Figure~\ref{fig:snls}. 
% Due to theoretical inequality constraints used to restrict the Galileon parameter space \citepalias{bib:neveu}, the number of degrees of freedom of the fits can not be computed exactly because parameters are then not independent. Note however that number fo degrees of freedom is dominated by the number of measurements whenever SNe~Ia enter the fits. In all tables in this paper, we thus provide only the fit global $\chi^2$'s.
Usually, one should fit and marginalise over the two nuisance parameters $\alpha$ and $\beta$, which describe the SN~Ia variability in stretch and colour \citep{bib:astier,bib:guy2010,bib:conley}. However, in \citetalias{bib:neveu} it was shown that for the Galileon model we can keep $\alpha$ and $\beta$ fixed to their marginalised values in the $\Lambda$CDM model. In this study, we thus took directly the fitted $\alpha$ and $\beta$ value from \citetalias{bib:neveu} for the SNLS3 sample and from \cite{bib:jla} for the JLA sample\footnote{The difference in $\alpha$ between the SNLS3 and JLA samples is due to different parametrisation of light-curves shapes: while for the SNLS3 sample, a stretch parameter $s$ is reported, the JLA sample uses the SALT2 $X_1$ parameter which is roughly $10\times(s-1)$.}.

Using the new JLA sample, we observe a $1\sigma$ increase of the best fit $\Omega_m^0$ value, as expected when considering the reported drift for the $\Lambda$CDM model in \cite{bib:jla}. Smaller changes are observed for the $\bar c_i$ parameters.

\subsection{Updated growth data computation}

From the Planck collaboration, we used the new value $\sigma_8^{\mathrm{Planck}}(z=0)=0.829\pm0.012$  \citep{bib:planck} to normalise our growth rate predictions, as we are able to remove the $\Lambda$CDM dependence for that observable. We recall that to compute $f\sigma_8(z)$ predictions in the Galileon model we assume that the value of $\sigma_8$ is equal in the Galileon and $\Lambda$CDM models at the decoupling redshift $z_*$. This is a reasonable assumption as we showed in our previous paper that the Galileon energy density is very subdominant at that time for most of the allowed set of parameters. The uncertainty on $\sigma_8^{\mathrm{Planck}}(z=0)$ was propagated in our error budget. We still used growth data with the same caveats as mentioned in \citetalias{bib:neveu} and \citetalias{bib:barreira13}, which we discuss further in Section~\ref{sec:disc}.

\begin{figure*}[hbtp]
\begin{center}
\epsfig{figure=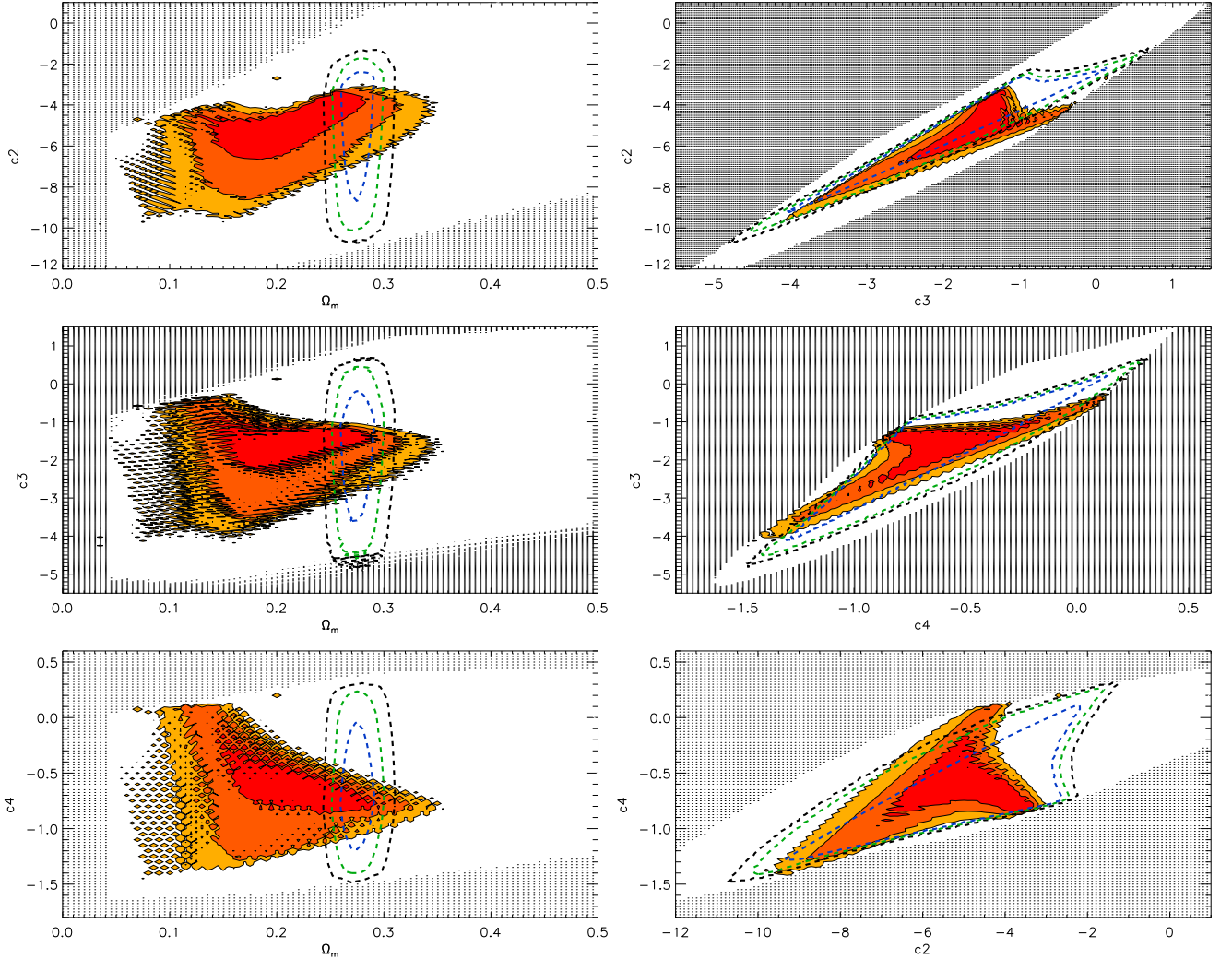, width=0.95\textwidth} 
\caption[]{Experimental constraints on the uncoupled Galileon model from growth data (red) and from JLA+WMAP9+BAO combined constraints (dashed). The filled dark, medium and light coloured contours enclose 68.3, 95.4 and 99.7\% of the probability, respectively. Dark dotted regions correspond to scenarios rejected by theoretical constraints.} 
\label{fig:gof_snlsbaowmap9}
\end{center}
\end{figure*}

\section{Uncoupled Galileon model}\label{sec:uncoupled}

In this section, we derived new experimental constraints on the $\bar c_i$ parameters of the uncoupled Galileon model following the same methodology as in \citetalias{bib:neveu} with updated data. 

\subsection{New experimental constraints}

\begin{table*}[htb]
\caption[]{Uncoupled Galileon model best-fit values from different data samples}
\label{tab:results}
\begin{center}
\begin{tabular}{cccccccc} \hline \hline \\ [-1ex]
Probe & $\Omega_m^0$ & $\bar c_2$ & $\bar c_3$ & $\bar c_4$ & $h$ & $\Omega_b^0h^2$ & $ \chi^2$ \\  [1ex] \hline \\ [-1ex]
%SNLS3 & $0.273^{+0.054}_{-0.042}$ & $-5.240^{+1.880}_{-2.802}$ & $-1.781^{+1.071}_{-1.426}$ & $-0.588^{+0.516}_{-0.348}$ & - & - & 420.1 \\ [1ex]  \hline \\ [-1ex]
%JLA  \\ [1ex]  \hline \\ [-1ex]
% avec bug sigma8 de wmap9: Growth & $0.257^{+0.049}_{-0.051}$ & $-4.687^{+0.822}_{-1.746}$ & $-1.519^{+0.315}_{-1.323}$ & $-0.606^{+0.264}_{-0.150}$ & - & - & 16.6\\ [1ex]  \hline \\ [-1ex] 
% best-fit avec sigma8 de Planck :
Growth data & $0.205^{+0.046}_{-0.046}$ & $-5.337^{+0.883}_{-1.293}$ & $-1.721^{+0.299}_{-0.732}$ & $-0.628^{+0.221}_{-0.189}$ & - & - & 20.1\\ [1ex]  \hline \\ [-1ex] 
BAO+WMAP9 & $0.274^{+0.015}_{-0.009}$ & $-5.467^{+1.962}_{-2.659}$ & $-1.896^{+0.996}_{-1.403}$ & $-0.622^{+0.462}_{-0.327}$ & 0.704 & 0.0226 & 0.7 \\ [1ex]  \hline \\ [-1ex] 
%BAO+WMAP9+H0  & $0.270^{+0.014}_{-0.009}$ & $-5.614^{+1.969}_{-2.650}$ & $-1.936^{+1.007}_{-1.404}$ & $-0.622^{+0.469}_{-0.335}$ & 0.714 & 0.0226 & 2.1 \\ [1ex]  \hline \\ [-1ex] 
%SNLS3+BAO+WMAP9+H0  & $0.270^{+0.014}_{-0.008}$ & $-5.585^{+1.955}_{-2.650}$ & $-1.925^{+1.000}_{-1.402}$ & $-0.622^{+0.469}_{-0.333}$ & 0.714 & 0.0226 & 423.2 \\ [1ex]  \hline \\ [-1ex] 
%chi2 = 0.68+421.39
SNLS3+BAO+WMAP9 & $0.274^{+0.014}_{-0.009}$ & $-5.463^{+1.952}_{-2.650}$ & $-1.892^{+0.992}_{-1.399}$ & $-0.621^{+0.462}_{-0.327}$ & 0.704 & 0.0226 & 422.1 \\ [1ex] \hline \\ [-1ex] 
%chi2 = 0.75 + 691.7
JLA+BAO+WMAP9 & $0.275^{+0.014}_{-0.009}$ & $-5.269^{+1.832}_{-2.726}$ & $-1.837^{+0.924}_{-1.408}$ & $-0.630^{+0.461}_{-0.304}$ & 0.701 & 0.0227 & 692.5 \\ [1ex] \hline  \hline \\ [-1ex] 

%chi2 = 421.16+5.96+22.73 à jour 
All (with SNLS3) & $0.270^{+0.013}_{-0.008}$ & $-4.315^{+0.525}_{-1.308}$ & $-1.568^{+0.201}_{-0.808}$ & $-0.759^{+0.101}_{-0.068}$ & 0.733 & 0.0220 & 449.9 \\ [1ex] \hline \\ [-1ex] 
All \citepalias{bib:neveu} & $0.271^{+0.013}_{-0.008}$ & $-4.352^{+0.518}_{-1.220}$ & $-1.597^{+0.203}_{-0.726}$ & $-0.771^{+0.098}_{-0.061}$ & 0.735 & 0.0220 & 450.4 \\ [1ex]  \hline \hline \\ [-1ex] 
% 421.13+6.03+22.68
% avec best-fit avec sigma8 de Planck et no H0 prior:
%chi2 = 691.67+5.8+23.4
All (with JLA) & $0.276^{+0.014}_{-0.009}$ & $-4.278^{+0.484}_{-1.097}$ & $-1.580^{+0.194}_{-0.597}$ & $-0.772^{+0.102}_{-0.058}$ & 0.726 & 0.0219 & 720.9\\ [1ex] \hline 
\end{tabular}
\tablefoot{SNLS3 and JLA with systematics included, $\alpha$ and $\beta$ fixed to their marginalised values. $h$ and $\Omega_b^0h^2$ have been minimized so no uncertainties are provided.}
\end{center}
\end{table*}

% à faire
\begin{table*}[htb]
\caption[]{Uncoupled Galileon model best-fit values compared with \citetalias{bib:barreira13} best-fit values, using SNLS3+WMAP9+BAO constraints.}
\label{tab:comparison}
\begin{center}
\begin{tabular}{ccccc} \hline \hline \\ [-1ex]
 & $\Omega_m^0$ & $\bar c_2/\bar c_3^{2/3}$ & $\bar c_4/\bar c_3^{4/3}$ & $\bar c_5/\bar c_3^{5/3}$\\  [1ex] \hline \\ [-1ex]
% no H0 prior: à jour
This work & $0.274^{+0.014}_{-0.009}$ & $-3.57^{+1.79}_{-2.47}$ & $-0.27^{+0.27}_{-0.24}$ & $0.12^{+0.23}_{-0.39}$  \\ [1ex]  \hline \\ [-1ex]
\citetalias{bib:barreira13}  & $0.273\pm0.010$ & $-4.04^{+0.35}_{-0.34}$ & $-0.171^{+0.035}_{-0.032}$ & $0.046^{+0.014}_{-0.017}$ \\ [1ex]  \hline \\ [-1ex]
\end{tabular}
\tablefoot{$\bar c_5$ is computed using equation (29) from \citetalias{bib:neveu}. The agreement between the best-fit values is better than 0.3$\sigma$ for all parameters.}
\end{center}
\end{table*}

Results using all probes are presented in Fig.~\ref{fig:gof_snlsbaowmap9}, and Table~\ref{tab:results}. With the SNLS3 data, the updated WMAP9 priors and the Planck $\sigma_8$ value improved only marginally the constraints and $\chi^2$ values of the Galileon model compared to our previous results using WMAP7 (see lines 5 and 6 in Table~\ref{tab:results}). %As can be checked in Table~\ref{tab:wmap}, although the WMAP9 power spectrum is more precise than the WMAP7 one, there is no gain on the distance prior uncertainties due to correlations between parameters. 
%For the growth data constraints, our previous result based on the WMAP7 value of $\sigma_8$ ($\sigma_8^{\mathrm{WMAP7}}(z=0)=0.811^{+0.030}_{-0.031}$) \textbf{showed a small} tension with distance measurements. %With the Planck $\sigma_8$ value, the agreement between the two sets of constraints is now marginally reduced. %However, the probability contours did not change significantly however, because even if we use more precise measurements, the overlap between contours is better and parameters are then less constrained. 
%Our best-fit values as well as the final $\chi^2$ are marginally improved compared to \citetalias{bib:neveu}. 
However, because this dataset is now more consistent with that used in \citetalias{bib:barreira13}, we can compare our two sets of results, obtained with different methodologies, which we do in the next section.

Finally, using the JLA sample does not improve the $\Omega_m^0$ uncertainty but decreases our uncertainties on the $\bar c_i$ Galileon parameters. This sample will be used to constrain the disformally coupled model in Section~\ref{sec:coupled}.%, as expected since the SNe Ia are a powerful probe to constrain dark energy parameters.}

\subsection{Comparison with \citetalias{bib:barreira13}}

\citetalias{bib:barreira13} recently provided constraints on the Galileon parameters, using the full WMAP9 CMB power spectrum whereas we used only distance priors. The rest of their dataset is identical to ours (except for one BAO measurement, which should not impact the result).

They used a different method to get rid of the degeneracies inherent to the original $c_i$ parametrisation, and derived constraints on $c_i/c_3^{i/3}$ ratios. Their method of computing the initial conditions is also different and more complex. Despite the different methodologies and parametrisations, the comparison of parameter ratios is possible as we have
\begin{equation}
\frac{\bar c_i}{\bar c_3^{i/3}}=\frac{c_i}{c_3^{i/3}}.
\end{equation}

This can give an interesting insight into the impact of the use of the full CMB power spectrum, and of different methodologies. Results using SNe~Ia, CMB and BAO data (no growth data) in both works are compared in Table~\ref{tab:comparison}. Both results are fully compatible even if our methodologies are different. Our $\Omega_m^0$ values and uncertainties are comparable, but the \citetalias{bib:barreira13} best-fit $c_i/c_3^{i/3}$ uncertainties are about ten times smaller than ours thanks to the use of the full CMB spectrum. 

%While the Galileon field is subdominant at decoupling redshift, interestingly far better constraints are obtained using the full CMB spectrum. 
This can be understood as follows. Distance priors are derived from the first acoustic peak only, which are measurements at the decoupling redshift where the Galileon field is subdominant. %Indeed, this information is contained in the low-$l$ region of the CMB power spectrum while distance priors only contain information on the first acoustic peak. 
But, as shown in \cite{bib:barreira}, in the Galileon theory the low-$l$ CMB power spectrum is very sensitive to the model parameters, because this part of the spectrum is affected by late-time dark energy through the integrated Sachs-Wolfe (ISW) effect. The Galileon model is thus severely constrained by this part of the spectrum, and can even provide a better fit to the CMB power spectrum than the $\Lambda$CDM model thanks to a better agreement in the low-$l$ region \citepalias{bib:barreira13}.

\begin{figure*}[hbtp]
\begin{center}
\epsfig{figure=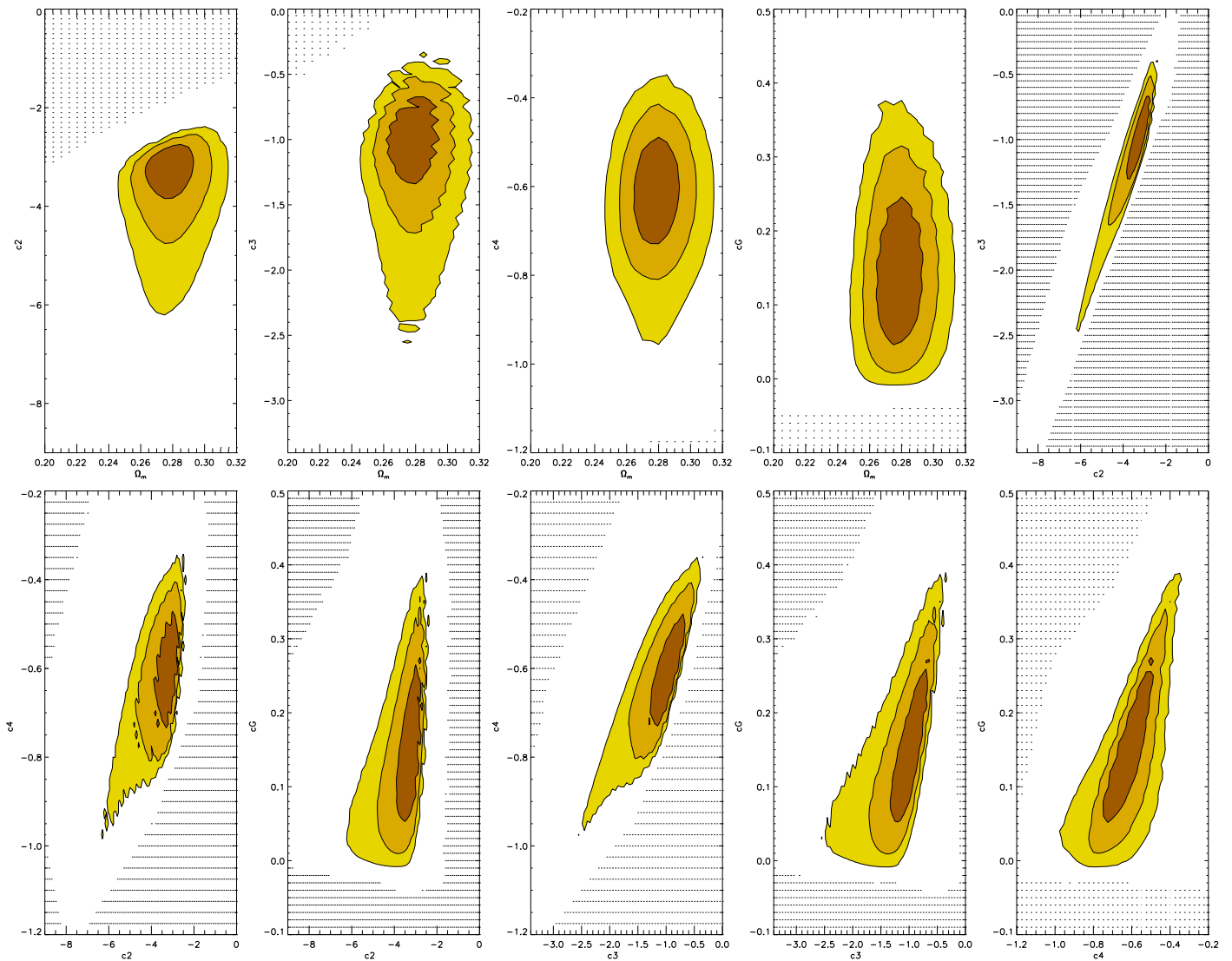, width=\textwidth} 
\caption[]{Combined constraints on the disformally coupled Galileon model from growth data combined with JLA+BAO+WMAP9 data. The filled dark, medium and light yellow contours enclose 68.3, 95.4 and 99.7\% of the probability, respectively. Dark dotted regions correspond to scenarios rejected by theoretical constraints.} 
\label{fig:gof_all_combined_cg}
\end{center}
\end{figure*}

\section{Galileon model disformally coupled to matter}\label{sec:coupled}

\subsection{Hypothesis}

In \cite{bib:appleby}, a disformal coupling $L_G$ between the matter and the Galileon field was proposed, motivated by extra-dimension considerations. $L_G$ introduces a new parameter to constrain, $c_G$. This kind of coupling naturally arises in the decoupling limit of massive gravity (see \cite{bib:deRhamMasGra2}). It also automatically arises when dealing with a fluctuating 3+1 brane in a $D=4+n$ dimensional bulk, when matter lives exclusively in the brane. This disformal coupling has already been studied in scalar field theories other than the Galileon model as reported in \cite{bib:brax12b}. 

In particular, \cite{bib:brax12b} and \cite{bib:andrews} showed that this kind of coupling has no gravitational effect on massive objects, and thus fulfils the Solar System gravity tests. However, it couples to photons and can play a role in gravitational lensing (see \cite{bib:wyman}).

In this work, as in \citetalias{bib:neveu}, we used the same cosmological and perturbation equations as in \cite{bib:appleby}, but the $c_G$  parameter was renormalised as
\begin{equation}
\bar c_G = c_G x_0^2,
\end{equation}
with $x_0$ defined in \citetalias{bib:neveu}.

Introducing this coupling did not require us to modify the methodology of \citetalias{bib:neveu}, but we had to assume that this coupling between matter and the Galileon field did not change the thermodynamics of the primordial plasma before decoupling. Indeed, if such a coupling is assumed, energy transfers should happen between the scalar field and the primordial plasma. In particular the disformal term leads to interactions between photons and the $\pi$ field. However, as shown in \citetalias{bib:neveu}, the uncoupled Galileon field is negligible during the radiation era in most scenarios, which limits the potential impact of the Galileon before the decoupling. We assumed that this is also the case here, in order to give a first glance at this interesting coupling.

\subsection{Results}

\begin{table*}[htb]
\caption[]{Disformally coupled Galileon model best-fit values from  growth rate measurements combined with JLA+BAO+WMAP9 data.}
\label{tab:resultscG}
\begin{center}
\begin{tabular}{ccccccccc} \hline \hline \\ [-1ex]
Probe & $\Omega_m^0$ & $\bar c_2$ & $\bar c_3$ & $\bar c_4$ & $\bar c_G$ & $h$ & $\Omega_b^0h^2$ & $ \chi^2$ \\  [1ex] \hline \\ [-1ex]
%SNLS3 & $0.270^{+0.031}_{-0.036}$ & $-4.688^{+1.568}_{-2.668}$ & $-1.097^{+1.045}_{-1.422}$ & $-0.232^{+0.531}_{-0.482}$ & $0.330^{+1.082}_{-0.238}$ & - & - & 421.2 \\ [1ex]  \hline \\ [-1ex]
%Growth & $0.278^{+0.027}_{-0.046}$ & $-3.767^{+0.756}_{-1.535}$ & $-1.124^{+0.273}_{-0.588}$ & $-0.538^{+0.136}_{-0.106}$ & $0.111^{+1.283}_{-0.065}$ & - & - & XXX\\ [1ex]  \hline \\ [-1ex] 
%BAO+WMAP9+H0  & $0.272^{+0.014}_{-0.009}$ & $-4.938^{+1.699}_{-2.581}$ & $-1.139^{+1.258}_{-1.415}$ & $-0.213^{+0.714}_{-0.480}$ & $0.389^{+1.080}_{-0.268}$ & 0.707 & 0.0228 & 2.4\\ [1ex]  \hline \\ [-1ex] 
%SNLS3+BAO+WMAP9+H0  & $0.272^{+0.014}_{-0.009}$ & $-4.829^{+1.654}_{-2.525}$ & $-1.057^{+1.290}_{-1.387}$ & $-0.180^{+0.709}_{-0.495}$ & $0.411^{+0.971}_{-0.284}$ & 0.707 & 0.0228 & 423.7\\ [1ex]  \hline 
%2.47+421.27
% with H0 prior and SNLS3
%All & $0.272^{+0.013}_{-0.008}$ & $-3.565^{+0.340}_{-0.769}$ & $-1.108^{+0.183}_{-0.278}$ & $-0.623^{+0.087}_{-0.080}$ & $0.127^{+0.086}_{-0.059}$ & 0.726 & 0.0221 & 444.5\\ [1ex]  \hline
% 421.27 + 4.63 + 18.64 = 444.5

% JLA+BAO+WMAP9
% chi2 = 691.57+1.59
JLA+BAO+WMAP9 & $0.282^{+0.015}_{-0.009}$ & $-4.811^{+1.427}_{-1.990}$ & $-1.525^{+0.637}_{-1.073}$ & $-0.531^{+0.209}_{-0.275}$ & $0.183^{+0.188}_{-0.133}$ & 0.689 & 0.0228 & 693.2 \\ [1ex]  \hline\\ [-1ex] 

% no H0 prior and JLA
All & $0.279^{+0.013}_{-0.008}$ & $-3.401^{+0.315}_{-0.565}$ & $-1.043^{+0.195}_{-0.252}$ & $-0.614^{+0.087}_{-0.076}$ & $0.147^{+0.077}_{-0.060}$ & 0.719 & 0.0220 & 714.8 \\ [1ex]  \hline
% chi2 = 691.62 + 19.04 + 4.14
\end{tabular}
%\tablefoot{SNLS3 with systematics included, $\alpha$ and $\beta$ fixed to their marginalised value. $h$ and $\Omega_b^0h^2$ have been minimized so no error bars are provided.}
\end{center}
\end{table*}

\begin{figure}[hbtp]
\begin{center}
\epsfig{figure=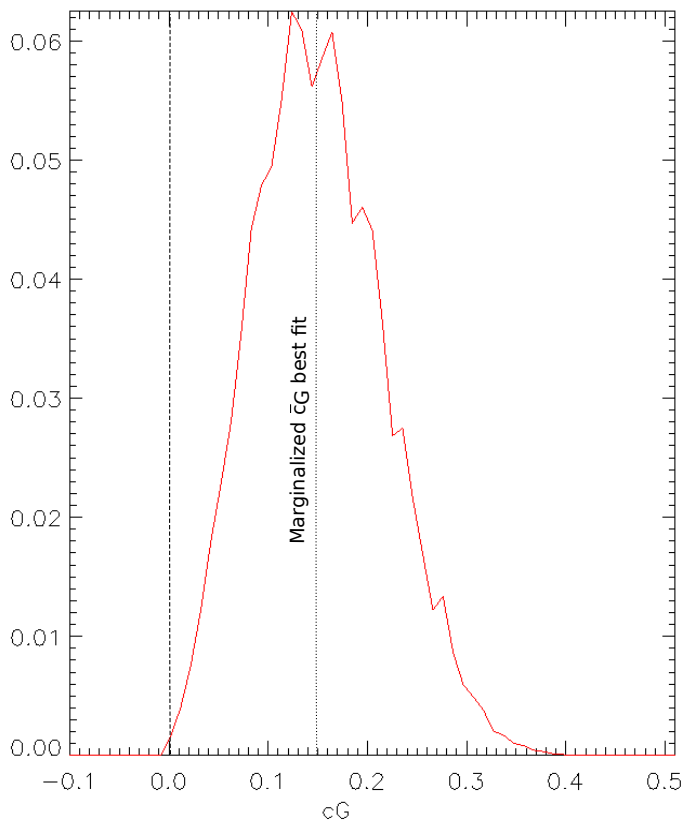, width=\columnwidth} 
\caption[]{$\bar c_G$ probability density function obtained using all datasets, and marginalising over all the other parameters.} 
\label{fig:gof_all_combined_cg_pdf_cgonly}
\end{center}
\end{figure}

The results using type Ia SNe, CMB, BAO measurements and growth data are presented in Fig.~\ref{fig:gof_all_combined_cg}, Fig.~\ref{fig:gof_all_combined_cg_pdf_cgonly}, and Table~\ref{tab:resultscG}.

As in the uncoupled case, the $\Omega_m^0$ best fit lies around 0.27. The $\bar c_2$ and $\bar c_3$ best fit values changed slightly, but are still compatible at one sigma with their best fit values in the uncoupled case, which is not the case for $\bar c_4$. As shown in Table~\ref{tab:resultscG}, $\bar c_G=0$ is excluded at the $2.5\sigma$ level, and as a consequence, the final $\chi^2$ is better than in the coupled case. 

The probability density function obtained for $\bar c_G$ (Fig.~\ref{fig:gof_all_combined_cg_pdf_cgonly}) shows clearly that a non-zero best fit value is preferred for this parameter, at the $2.5\sigma$ level. Note that when using distances only, this result still holds but at the $1.4\sigma$ level. The disformally coupled Galileon model appears thus in better agreement with data than the uncoupled model. %\textbf{However, this constraint can not be directly compared to results from \cite{bib:brax14} and \cite{bib:monophoton} due to the new parametrisation with $\bar c_G$.} % This could have been expected because we added one more degree of freedom to the model. XXXXXXXXXXX But as the best-fit value obtained for this parameter excluded at the $2.2\sigma$ level the previously obtained best-fit value, the introduction of this parameter was not superfluous and improved the model. XXXXXXXXXXXXXXXXXXXXXXXXXXXXXXXXX 
This may support an extra-dimension origin for the Galileon theory as this coupling is unavoidable in such constructions. 

\subsection{Implications beyond cosmology}

The parameter space explored in this work was defined by theoretical conditions that ensure that our cosmological solution is free of ghosts and instability problems (see section 2.5 of \citetalias{bib:neveu}).

Recently, \cite{bib:berezhiani} and \cite{bib:koyama} pointed out that some Galileon-like theories may lead to ghost instabilities in solutions inside massive objects, when a disformal coupling is involved. One should note, however, that our model does not fall exactly in their discussion. However, following a reasoning similar to that in \cite{bib:berezhiani}, we checked that our model may avoid the ghost problem thanks to a compensation between the fifth Lagrangian term $L_5$ and $L_G$. Thus, our best-fit result is likely to be valid also inside massive objects, but deriving an explicit no-ghost condition to ensure that this is indeed the case is beyond the scope of this paper, since we restricted to cosmological solutions only.

%Recently, \cite{bib:berezhiani} and \cite{bib:koyama} expressed concerns about solutions of Galileon-like theories inside massive objects when a disformal coupling is involved. They showed that these solutions may exhibit instability problems in some of these theories. The Galileon model studied in this paper was not part of their discussion. However, following a reasoning similar to that in \cite{bib:berezhiani}, we checked that our model may be free of this sort of ghost problems thanks to a compensation between the fifth Lagrangian term $L_5$ and $L_G$. Thus, our best fit result is likely to be valid also inside massive objects, but deriving an explicit no-ghost condition to ensure that this is indeed the case is beyond the scope of this paper, since we restricted to cosmological solutions only.

\section{Discussion}\label{sec:disc}

\begin{figure*}[hbtp]
\begin{center}
\epsfig{figure=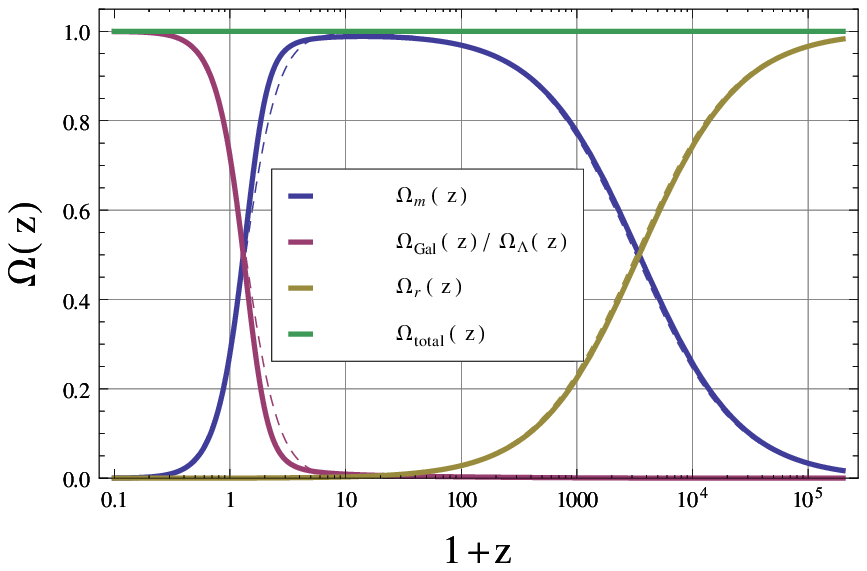, width=1.05\columnwidth} 
\epsfig{figure=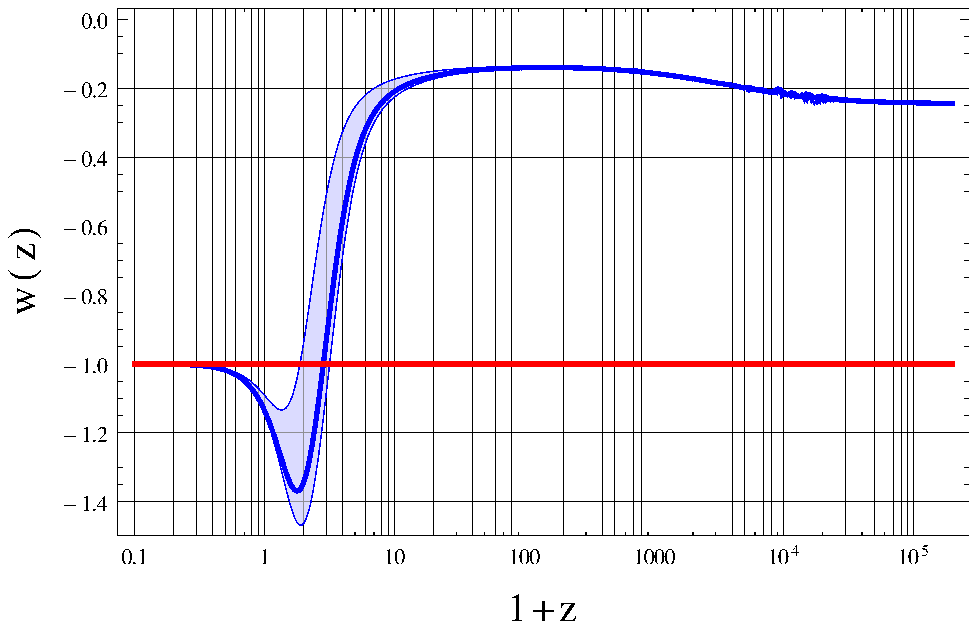, width=0.95\columnwidth} 
\caption[]{Evolution of the $\Omega_i(z)$ (left) and of $w(z)$ (right, solid curve) for the best-fit disformally coupled Galileon model from all data (last row of Table~\ref{tab:resultscG}). In the left plot, dashed lines correspond to our $\Lambda$CDM best-fit values. Differences in the radiation era are only due to different best-fit values of $h$. In the right plot, the shaded area was obtained varying the $\bar c_i$ parameters within their $1\sigma$ uncertainties, taking their correlations into account.} 
\label{fig:omega}
\end{center}
\end{figure*}
\subsection{Non-linearities in the Galileon model}

In \citetalias{bib:neveu}, we provided some caveats about the use of growth data when comparing with predictions of a linearised Galileon model. In particular, we stressed that:
\begin{itemize}
\item non-linear models of growth of structures were used to extract measurements from data, whereas here only the linear perturbation theory was used to describe matter perturbation evolution,
\item the non-linearities of the Galileon field itself were also neglected in the perturbed equations while they play a key role in the Vainshtein effect, in particular to restore general relativity close to massive objects.
\end{itemize}
In particular for the latter point, below a certain distance from a massive object, called the Vainshtein radius, Galileon gravity is supposed to vanish with respect to general relativity. At small scales, the growth of structures is then identical in both models. But the Vainshtein radius depends on the Galileon parameters and on the massive object properties, and hence it is difficult to know which scales are affected by Galileon gravity in general. 

However, recent progress on the inclusion of non-linearities (both from the Galileon field and from matter evolution) has been made using N-body simulations of structure evolution in Galileon gravity \citep{bib:barreira13b,bib:li13,bib:barreira13c}. Matter power spectra were computed at the non-linear level for different Galileon models. In \cite{bib:barreira13b}, the cubic Galileon model (i.e. with $c_4=c_5=0$) was studied. It was shown that, for scales $k$ between $0.1$ and $0.4\ h$Mpc$^{-1}$ (the ones encompassed in our growth data measurements), the deviation between a linear and a non-linear Galileon model is at most of $\approx 5$\%, and only at redshifts below 0.2.  The authors then studied a quartic Galileon model in \cite{bib:li13}. Important deviations appear between the linear and the non-linear model for the velocity divergence power spectrum (the one of interest for redshift space distortions), but the authors recommended more precise simulations before drawing firm conclusions on that point. Finally, \cite{bib:barreira13c} showed that the quintic Galileon model (the one we are using) has non-physical solutions in high matter density regions and that prevented them from doing simulations in that case. According to the authors, these non-physical solutions are either inherent to the Galileon model itself, or due to the approximations made in their non-linear computations. Further work has to be done on the non-linear modelling of the quintic Galileon gravity to make reliable predictions for the growth of structures.

\subsection{Comparison with the $\Lambda$CDM model}

% à jour
\begin{figure}[hbtp]
\begin{center}
\epsfig{figure=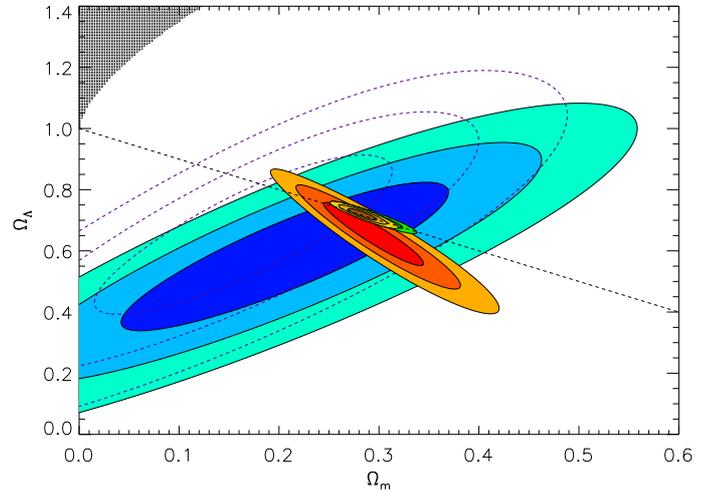, width=\columnwidth} 
\caption{Experimental constraints on the $\Lambda$CDM model from JLA data (blue), growth  data (red), BAO+WMAP9 data (green), and all data combined (yellow). Purple dashed contours stand for the $\Lambda$CDM constraints using SNLS3 data only (combining SIFTO and SALT2 supernova parameters). The black dashed line indicates the flatness condition $\Omega_m+\Omega_\Lambda=1$.} 
\label{fig:lcdm}
\end{center}
\end{figure}

% chi2_LCDM pour JLA+BAO+WMAP9 = 691.55+1.43=693.0
% chi2_LCDM pour growth+JLA+BAO+WMAP9 = 691.58+1.40+12.49
% CHI2_LCDM pour growth+SNLS+BAO+WMAP9=421.48 + 6.00 + 11.36 = 438.9
\begin{table}[htb]
\caption[]{$\chi^2$s at marginalised values for different models and different datasets.}
\label{tab:chi2}
\begin{center}
\begin{tabular}{ccc} \hline \hline \\ [-1ex]
 Probes  &  JLA+BAO+WMAP9 & All \\  [1ex] \hline \\ [-1ex]
$\Lambda$CDM & 693.0 & 705.5 \\ [1ex]  \hline \\ [-1ex]
Uncoupled Gal. & 692.5 & 720.9 \\ [1ex]  \hline \\ [-1ex]
Coupled Gal. & 693.2 & 714.8 \\ [1ex]  \hline \\ [-1ex]
\end{tabular}
\end{center}
\end{table}

The best-fit coupled Galileon scenario still mimics a $\Lambda$CDM model with the three periods of radiation, matter, and dark energy domination, with an evolving dark energy equation of state parameter $w(z)$ (see Fig.~\ref{fig:omega}). This feature is not different from what we obtained in the uncoupled case.

Table~\ref{tab:chi2} presents the $\chi^2$ values of the two Galileon models and the $\Lambda$CDM one (see also Fig.~\ref{fig:lcdm}). With only distances, the three models reach the same level of agreement with data. When adding growth data, the increase in $\chi^2$ is higher for the Galileon models, as a result of a higher tension between growth and distance probes (see also Fig.~\ref{fig:gof_snlsbaowmap9} and \ref{fig:lcdm}). But the difference in $\chi^2$ with respect to the $\Lambda$CDM model is not stringent. As in our previous work, we can conclude that the Galileon model is not significantly disfavoured by current data compared to the $\Lambda$CDM model, and is a good alternative to model dark energy.

\section{Conclusion}\label{sec:concl}

We have compared the uncoupled and disformally coupled Galileon models to the most recent cosmological data, using the methodology from our previous work \citepalias{bib:neveu}. 
An update of the uncoupled Galileon model experimental constraints using WMAP9 $\left\lbrace l_a,R,z_* \right\rbrace$ constraints was derived jointly with the new JLA SN~Ia sample, BAO measurements, and growth data with the Alcock-Paczynski effect taken into account. The $\sigma_8(z=0)$ value used to compute the growth of structure observable was also updated to the \cite{bib:planck} value. The JLA sample allowed us to derive better constraints on the $\bar c_i$ parameters. When we kept the SNLS3 sample, our constraints did not change significantly, but led to an interesting comparison with the Galileon best-fit values published in \citetalias{bib:barreira13}. They used the full WMAP9 power spectrum to derive their constraints, and thus brought tighter constraints, but both best-fit scenarios agree. This validates both methodologies despite their differences. As expected,  WMAP9 distance priors are less constraining than the full CMB spectrum but provide a simpler and faster way to derive constraints on the Galileon model.  

We provided the first experimental constraints on the disformal coupling parameter in the framework of the Galileon model. This coupling between matter and the Galileon field is natural when building the theory from massive gravity or extra-dimension considerations. Our final $\chi^2$'s are comparable to the one obtained for the $\Lambda$CDM model. Galileon theories are thus competitive to explain the nature of dark energy. 
We also showed that a null disformal coupling to matter is excluded at the $2.5\sigma$ level when using growth data, and at the 1.4$\sigma$ level when using distances only. %, and thus the coupled Galileon model fits data better than the uncoupled model. 
This gives some interesting clues, from experimental data, on the possible extra-dimension origin of Galileon theories. 

Better constraints would be possible including the ISW effect as shown in \citetalias{bib:barreira13}. The galaxy velocity field could also be a decisive probe to test modified gravity theories, as advocated in \citep{bib:zu,bib:hellwing}. However, this probe would require to have a correct modelling of the Galileon model non-linearities. %In the light of Fig.\ref{fig:h_ratios}, we also propose that precise additionnal distance measurements probing the dark energy dominated era and the matter dominated era transition could have a significant impact (between redshifts 0 and 3). Indeed, in this range of redshift we observe a significative deviation between our $\Lambda$CDM and Galileon best fit scenarios. In particular, precise distance measurements at redshift above 2 using BAOs or Lyman-$\alpha$ forests from the eBOSS experiment are compelling \citep{XXXXXXXXXXX}.
Interestingly, the disformal coupling couples to light and thus can have an impact on gravitational lensing \citep{bib:wyman}. Lensing experiments such as LSST \citep{bib:lsst} %\footnote{http://www.lsst.org/lsst/}
or the future satellite Euclid \citep{bib:euclid}, or laboratory tests with light shining through a wall experiments \citep{bib:brax12b} can provide more data to constrain this interesting coupling. CMB spectral distortion studies \citep{bib:brax13} will also give further insight into the braneworld origin of the Galileon theory.

%\begin{acknowledgements} 
%
%We thank Philippe Brax for introducing us with the Galileon theory, and Christos Charmoussis, C\'edric Deffayet and Jean-Baptiste Melin for fruitful discussions about the Galileon model. We also thank Chris Blake for useful advice on the use of the WiggleZ measurements. The work of Eugeny Babichev was supported in part by grant FQXi-MGA-1209 from the Foundational Questions Institute.
%
%\end{acknowledgements} 

\end{document}